\documentclass[12pt,preprint]{aastex}
\usepackage{t1enc}
\usepackage{graphicx}
\usepackage{epstopdf}
\usepackage{natbib}
\usepackage{lscape}
\usepackage{booktabs}
\newcommand{\ammonia}{NH$_{3}$}
\newcommand{\hcom}{HCO$^+$}

\newcommand{\mum}{$\mu$m}
\newcommand{\kms}{km~s$^{-1}$}

\newcommand{\msun}{$M_{\odot}$}

\newcommand{\cmt}{cm$^{-3}$}
\newcommand{\nhtd}{NH$_2$D}
\newcommand{\hntc}{HN$^{13}$C}
\newcommand{\jpb}   {$\rm Jy~beam^{-1}$}
\newcommand{\vel}{$v_\mathrm{LSR}$}
\newcommand{\Trot}{$T_\mathrm{rot}$}
\newcommand{\cmd} {cm$^{-2}$}
\newcommand{\Tex}   {$T_\mathrm{ex}$}
\newcommand{\gap}%
{\raisebox{-0.5ex}{$\stackrel{\scriptstyle >}{\scriptstyle \sim}$}}

\shorttitle{Study of the dense core ahead of HH 80N}
\shortauthors{Masqu\'e et al.}

\begin{document}

\title{Interferometric observations of nitrogen-bearing molecular species in the star-forming core ahead of HH 80N}

\author{
Josep M. Masqu\'e\altaffilmark{1,2},
Josep M. Girart\altaffilmark{3},
Guillem Anglada\altaffilmark{2},
Mayra Osorio\altaffilmark{2},
Robert Estalella\altaffilmark{1},
\and
Maria T. Beltr\'an\altaffilmark{4}
 }

\altaffiltext{1}{Departament d'Astronomia i Meteorologia, Universitat de Barcelona, 
Mart\'i i Franqu\`es 1, 08028 Barcelona, Catalunya, Spain}

\altaffiltext{2}{Instituto de Astrof\'isica de Andaluc\'ia, CSIC, Camino Bajo de Hu\'etor 50, E-18008 Granada, Spain}

\altaffiltext{3}{Institut de Ci\`encies de l'Espai, (CSIC-IEEC), 
Campus UAB, Facultat de Ci\`encies, Torre C5 - parell 2, 
08193 Bellaterra, Catalunya, Spain}

\altaffiltext{4}{INAF - Osservatorio Astrofisico di Arcetri, Largo E. Fermi 5, 50125 Firenze, Italy}

\begin{abstract}

We present VLA \ammonia\ and PdBI \nhtd\ and \hntc\ observations of the star forming core ahead of HH 80N, the optically obscured northern counterpart of the Herbig-Haro objects HH 80/81. The main goal is to determine the kinematical information of the high density regions of the core ($n$~\gap~$10^5$~\cmt), missed in previous works due to the depletion of the species observed (e.g. CS). The obtained maps show different kinematical signatures between the eastern and western parts of the core, suggesting a possible dynamical interaction of the core with the HH 80/81/80N outflow. The analysis of the Position-Velocity (PV) plots of these species rules out a previous interpretation of having a molecular ring-like structure of $6 \times 10^4$~AU of radius traced by CS infalling onto a central protostar found in the core (IRS1). High degree of \ammonia\ deuteration, with respect to the central part of the core harboring IRS1, is derived in the eastern part, where a dust condensation (SE) is located. This deuteration trend of \ammonia\ suggests that SE is in a prestellar evolutionary stage, earlier than that of the IRS1. Since SE is the closest condensation to the HH 80N/81/80N outflow, in case of having outflow-core dynamical interaction, it should be perturbed first and be the most evolved condensation in the core. Therefore, the derived evolutionary sequence for SE and IRS1 makes the outflow triggered star formation on IRS1 unlikely.

\end{abstract}

\section{Introduction}

Star formation can be considered a process where there are almost simultaneous infall and outflow motions \citep[e.g.,][]{evans1999}. 
The outflow phenomenon is very well studied because it has numerous observational signpost associated: Herbig-Haro objects, optical jets, thermal radio jets, molecular outflows (e.g. see \citealt{anglada1996a} and \citealt{richer2000}). On the other hand, the study of the gas kinematics in the region of protostellar collapse is often hindered by the complexity of the gas motions resulting from a combination of rotation, turbulence and/or stellar outflows \citep[e.g.,][]{belloche2002}. In addition, a number of molecular species freeze out onto icy dust mantles at temperatures $\leq$ 20~K and densities \gap~10$^5$~cm$^{-3}$ \citep{bergin1995, aikawa2001, jorgensen2004}, corresponding to the properties of the gas in the inner regions of the core where the collapse may take place. Therefore, the selection
of a molecule that has little or no depletion, observed with an instrument with high angular resolution such as an interferometer, is a request to trace the inner and compact regions of the core. Detailed studies show that, once the contraction of a dense core starts, the increasing central density yields to chemical inhomogeneities \citep{rawlings1992,bergin1997}. As the core evolves,   
while molecules such as CS or CO deplete first, nitrogen-bearing molecules can remain on the gas phase at densities higher than 10$^5$~cm$^{-3}$ \citep{tafalla2002, tafalla2004, aikawa2003}. Thus, N-bearing molecules tend to trace the dense inner parts of the core better than other species do. Furthermore, a second-order effect induced by the CO freeze-out is a sharp rise of the abundance in the gas phase of deuterated species, which are efficiently produced at low temperatures.
In this sense, following the same trend as N-bearing molecules, deuterated molecules are found to be key probes for extremely cold ($\simeq$10~K) and dense (10$^5$--10$^6$~\cmt) 
gas \citep[e.g.][]{caselli2002b}. Therefore, they are ideal to map the central part of the core at the onset of collapse. 

HH 80N is the optically obscured northern head of the HH 80/81/80N system, the largest collimated radio-jet associated with an YSO \citep{rodriguez1980,marti1993}. It is located in the GGD 27 region at 1.7 kpc of distance \citep{rodriguez1980}. Ahead of HH 80N, there is a dense core of 0.3~pc in size (hereafter HH 80N core) located $\sim0.3$~pc from the HH object. The core was first detected in ammonia \citep{girart1994} and afterward in other molecular species \citep{girart1998, girart2001, masque2009}. A mass of roughly 20~\msun\ and an average rotational temperature of $\sim17$~K was estimated by \citet{girart1994} from the ammonia emission of the core. Molecular observations suggest that the chemistry of the section of the core facing HH 80N is being altered by the UV photons coming
from the HH object \citep{girart1998, masque2009}. This suggests that the association core-HH object is real and not an effect of the sky projection. 

\citet{girart2001} found evidence of a bipolar CO outflow centered near the peak of the HH 80N molecular core, suggesting the presence of an embedded protostar (IRS1, \citealt{masque2011}), which is detected by
Spitzer \citep{masque2009}. In addition, based on the CS PV plots appearance, 
\citet{girart2001} and \citet{masque2009} interpreted the morphology and kinematics of the HH 80N core as a ring-like
structure of $6 \times 10^4$~AU, falling onto the central embedded object. The ring-like structure seen with molecular tracers observed with BIMA would not be a real structure but the result of a strong molecular depletion at the inner region of the core \citep{girart2001}. 


However, high angular resolution observations obtained with the Plateau de Bure Interferometer (PdBI), have revealed that the HH 80N core is composed of several dusty condensations suggesting a possible fragmentation for this core \citep{masque2011}. The main condensation is associated with the embedded protostellar object, IRS1, while another condensation (SE) has prestellar nature and is likely close to collapse. 
The IRS1 condensation was modelled as a slowly rotating collapsing envelope of 20~\msun\ of mass, with an infalling region of 1.5$ \times 10^4$~AU of radius, and the rest of envelope outside this radius being static. In order to explain the kinematical signatures detected with CS in the HH 80N region, \citet{masque2011} proposed that this molecule traces diffuse gas within the HH 80N core surrounding the SE and IRS1 condensations. This gas component, which has an estimated mass of $\sim10$~\msun, would be gravitationally unbound from IRS1, with its kinematics possibly affected by the HH 80/81/80N outflow.

In order to observationally confirm the scenario of \citet{masque2011}, we have carried out high angular resolution observations of N-bearing and deuterated
molecules to reveal the kinematic properties of the high density  
molecular gas ($n$~\gap~$10^5$~\cmt). This dense gas is found in regions of the core missed in  
our previous work due to the depletion of the species observed (e.g. CS). Our main goal is to disentangle the
gas motions belonging to the protostellar collapse of IRS1 from those
associated to large scale kinematics of the diffuse gas traced by CS. In \S~2 we describe the data reduction and present the results. In \S~3 we analyze the kinematics and the deuteration enrichment of the HH 80N core and discuss the  possibility of having induced star formation in the core. Finally, our conclusions are summarized in \S~4.

\section{Data reduction and results}

We carried out VLA \ammonia~(1,1) and \ammonia~(2,2) and PdBI \nhtd~(1$_{1,1}$--1$_{0,1}$) and \hntc~(1--0) observations of the HH 80N region. The observational setup was described in \citet{masque2011}, where we report the integrated \ammonia~(1,1) emission map and the PdBI 3.5~mm continuum emission results. The VLA ammonia maps presented here were obtained with natural weighting and using a Gaussian
taper of 35~k$\lambda$, which gives a beam size of $6\farcs4 \times 4\farcs4$ (P.A. = 17.2$\arcdeg$)  for
the \ammonia~(1,1) transition maps and $6.2\arcsec \times 4.3\arcsec$ (P.A. = 18.3$\arcdeg$) for the
\ammonia~(2,2) transition maps.  The {\em rms} noise level is 4~mJy~beam$^{-1}$ and 5~mJy~beam$^{-1}$ {\em per} channel for the \ammonia~(1,1) and \ammonia~(2,2) maps, respectively. The PdBI \nhtd~(1$_{1,1}$--1$_{0,1}$) and HN$^{13}$C~(1--0) channel maps were obtained using MAPPING with
natural weighting, that gives a synthesized beam (HPBW) of $7\farcs0 \times 2\farcs9$ (P.A. = $11\arcdeg$) at 86~GHz. The  {\em rms} noise level of the
channel maps are  20~mJy~beam$^{-1}$ and
23~mJy~beam$^{-1}$ {\em per} channel for \nhtd~(1$_{1,1}$--1$_{0,1}$) and \hntc~(1--0), respectively. A summary of the parameters of the VLA and PdBI observations is given in Table~\ref{liniesobservades}.

\begin{figure}[htbp]
\centering
\resizebox{\textwidth}{!}{\includegraphics{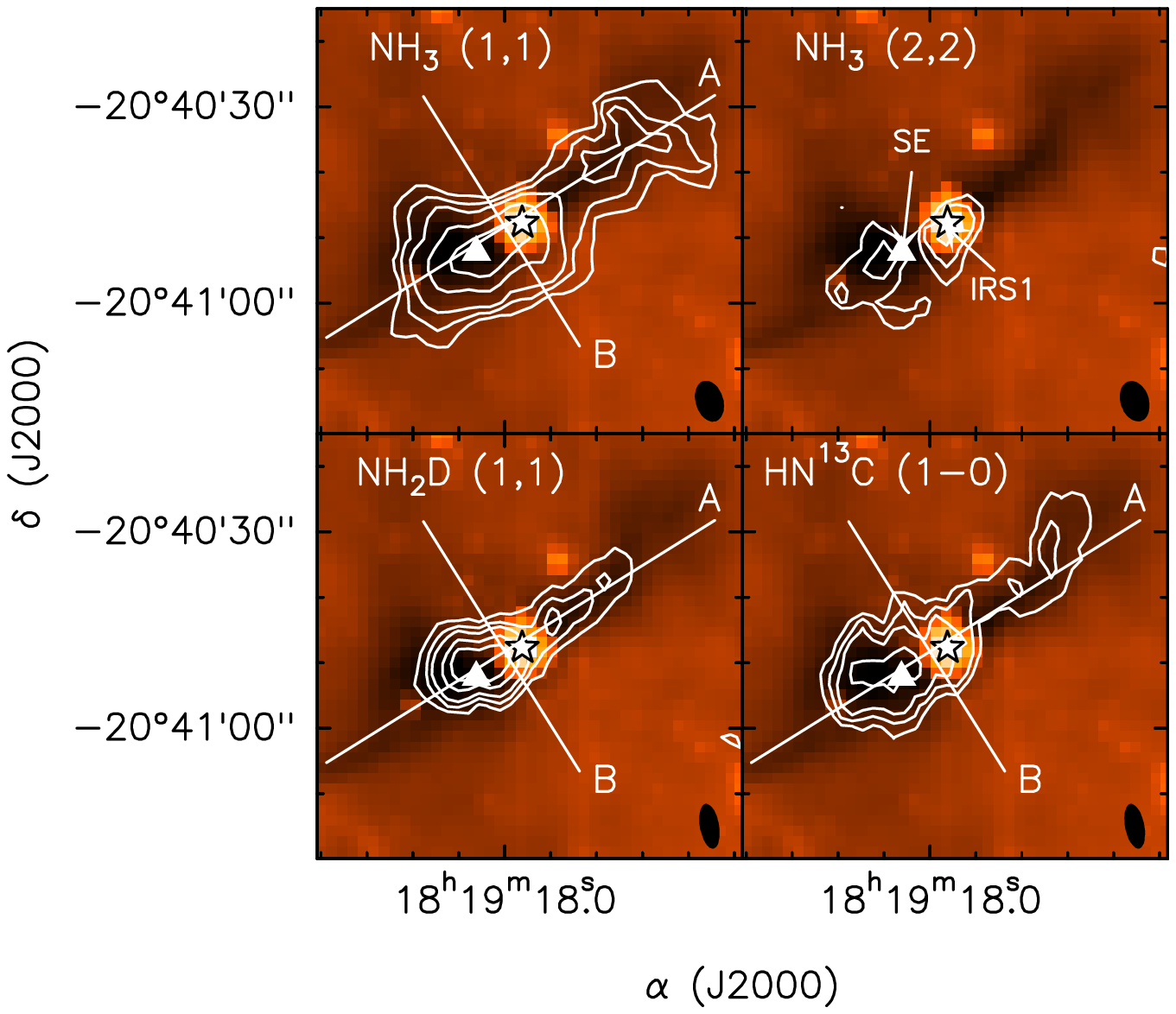}}
\figcaption{
Superposition of the Zero-order moment maps (integrated emission;
white contours) of the N-bearing species in the HH 80N core,
observed with VLA and PdBI, overlaid with the Spitzer 8~\mum\ image (color scale). The integrated emission of \ammonia~(1,1) and \ammonia~(2,2) corresponds to the main
line of the transition.  The contour levels are 1, 2, 3, 5, 7, 10 and 15 times  
 7~m\jpb~\kms\ (\ammonia~(1,1)), 3.5~m\jpb~\kms\ (\ammonia~(2,2)), and 30~m\jpb~\kms\ (\hntc\ and \nhtd). The \ammonia~(1,1) map was reported in \citet{masque2011}.
The solid lines show the cuts for the A and B
direction PV plots presented in Figs.~\ref{pvplots_n} and \ref{pvplots_cs}. The star and triangle mark the positions of IRS1 and SE, respectively. The beam is shown in the bottom right
corner of each panel. \label{ammonia_integrated}}
\end{figure}

We detected all the species listed in Table~\ref{liniesobservades} in the 10.5 to 12.3~\kms\ velocity range. The integrated intensity maps are
shown in Figure~\ref{ammonia_integrated}. The morphology of the integrated emission of \ammonia, \nhtd\ and \hntc\ is elongated in the SE-NW direction, being consistent with the elongation displayed by the species observed with BIMA \citep[mainly CS, \hcom\ and SO,][]{masque2009}. Also, the elongation of N-bearing species follows approximately the dark lane seen in the $Spitzer$ 8~\mum\ image. \ammonia~(2,2) is weaker than the other lines of Table~\ref{liniesobservades} and only the two main condensations, IRS1 and SE, have emission associated.

Despite having a similar shape, the emission of the N-bearing species presented in this work appears more
compact than the emission of the species observed in \citet{masque2009}. \ammonia~(1,1) is the
most extended having a FHWM angular size of 50\arcsec $\times$ 15\arcsec\ ($0.41 \times 0.12$~pc).
\hntc\ and \nhtd\ present smaller FHWM sizes of 25\arcsec $\times$ 10\arcsec\ ($0.21 \times 0.08$~pc). These sizes are clearly smaller than, for example, the size of 65\arcsec $\times$ 25\arcsec\ ($0.54 \times 0.21$~pc) derived for CS. The maps of Fig.~\ref{ammonia_integrated} also reveal two main trends: \ammonia~(2,2) peaks at the IRS1 position while \ammonia~(1,1), \nhtd\ and \hntc\ are brightest
towards the SE condensation.

\begin{landscape}
\begin{figure}[htbp]
 \centering
\resizebox{1.4\textwidth}{!}{\includegraphics[angle=270]{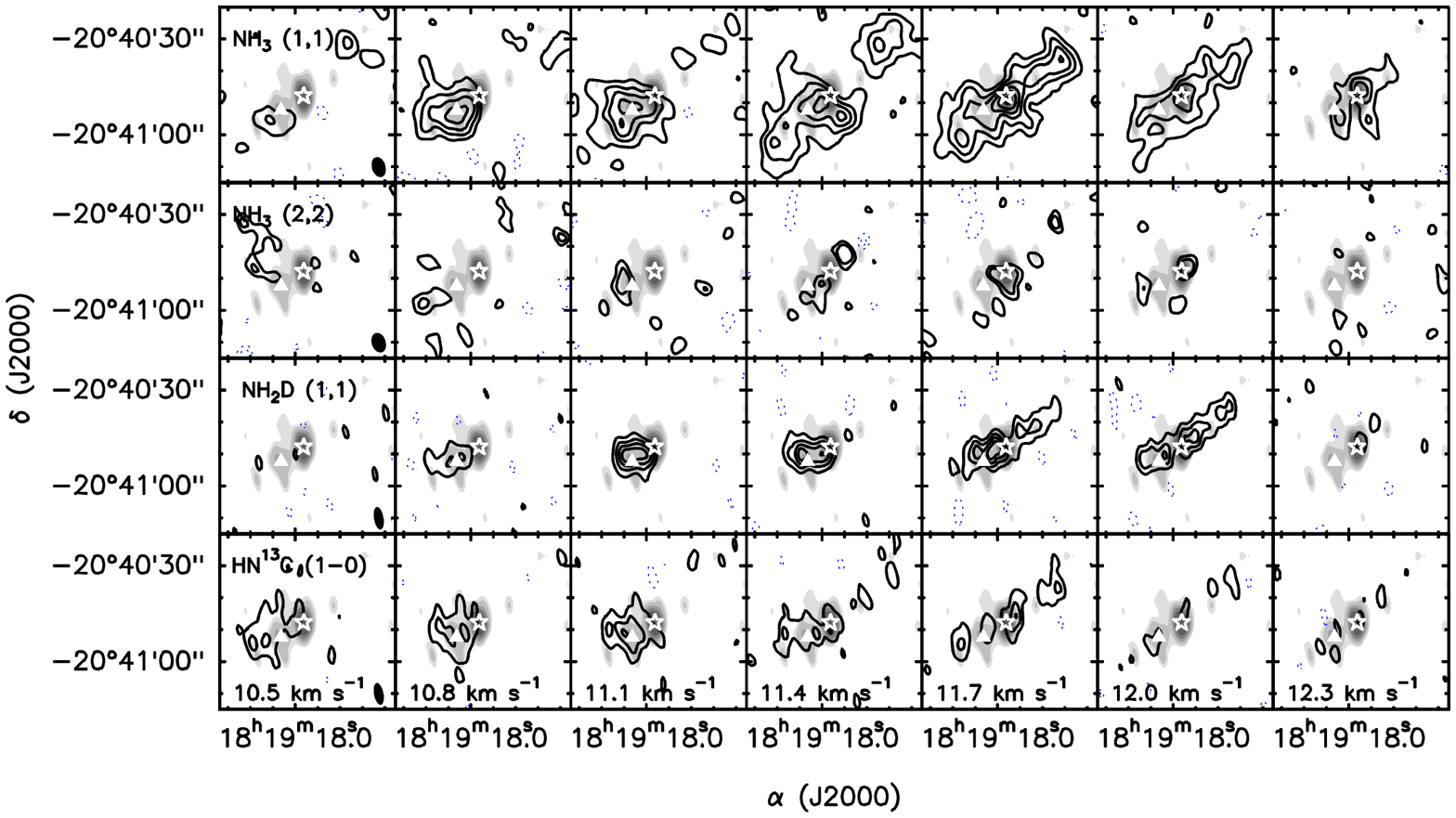}}
\caption{Contour maps of the velocity channels from 10.5~\kms\ to 12.3~\kms\ of the \hntc~(1--0), \nhtd~(1,1), \ammonia(2,2) and \ammonia (1,1) lines (from top to bottom), with a channel width of
0.3~\kms, superimposed over the 3.5~mm continuum PdBI map of  \citet{masque2011} (gray scale). The contour levels
 are  -3, 3, 6, 9, 12, 15 and 18 times the rms noise level for each species (see Table~\ref{liniesobservades}).
For the \ammonia~(2,2) line, the contours -2 and 2 times the rms noise level are also shown. The symbols are the same as in Fig.~\ref{ammonia_integrated}.
The beam is shown in the bottom right corner of the panels of the first column.
\label{chan_alta}}
\end{figure}
\end{landscape}

\begin{landscape}
\begin{figure}[htbp] 
\centering
\resizebox{1.5\textwidth}{!}{\includegraphics{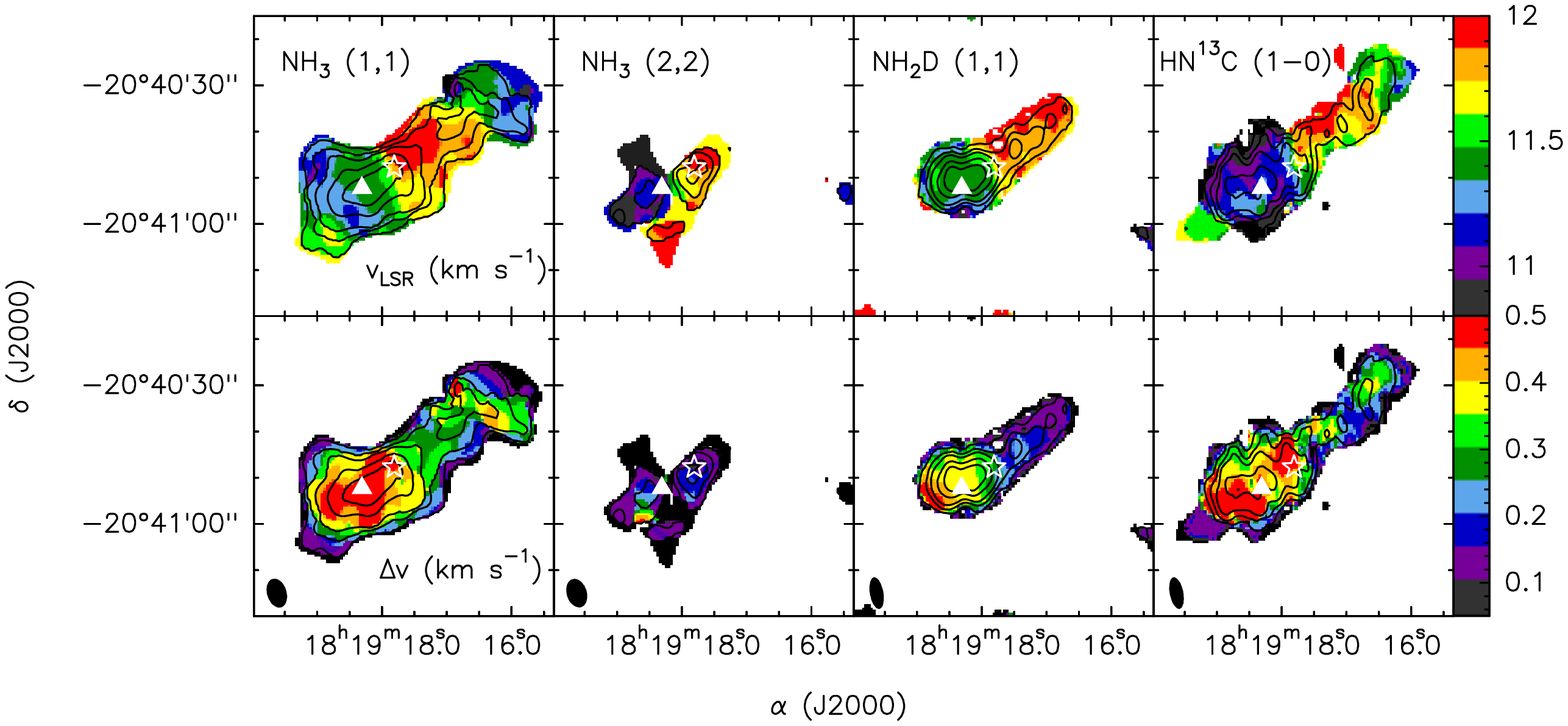}}
\caption{
Superposition of the contour maps of the zero-order moment (integrated emission) with the color image maps of first-order
(top panels) and second-order (bottom panels) moments of the \ammonia~(1,1), \ammonia (2,2), \hntc~(1--0) and \nhtd~(1,1) lines (from left to right). The moment maps were obtained with the 10.1--12.8~\kms\ \vel\ range for \ammonia~(1,1), 10.7--12.2~\kms\  for \ammonia (2,2) and \hntc, and 10.4--12.8~\kms\ for \nhtd. Contour levels of the zero-order moment maps and symbols are the same as in Fig.~\ref{ammonia_integrated}. 
The beam is shown in the bottom left corner of the panels of the bottom row.
\label{moments}}
\end{figure}
\end{landscape}

Figure~\ref{chan_alta} shows the channel maps over the 10.5 to 12.3~\kms\ velocity
range, where most of the emission of the N-bearing species is detected, superposed on the
continuum 3.5~mm PdBI map of \citet{masque2011}. The blue-shifted emission (\vel~from 10.5 to 11.4~\kms) of the
N-bearing molecules is mainly distributed to the southeast of IRS1, in the vicinity of SE. 
The rest of the emission (\vel~from 11.4 to 12.3~\kms) is widely distributed along all the core with a size similar to that of the dark lane seen in the $Spitzer$ 8~\mum\ image and beyond the region traced by the 3.5~mm continuum emission. Among the lines of the N-bearing species presented here, the \ammonia~(1,1) line has the most extended emission. Besides, \hntc\ and \nhtd\ exhibit differences between them: while most \nhtd\
emission appears concentrated towards the SE condensation, \hntc\ shows discrete peaks along SE-NW direction, one of them coinciding
with IRS1. In Figure~\ref{moments} we show the maps of the first-order moment
(top panel) and second-order moment (bottom panel) superposed on the zero-order
moment. This figure better illustrates the differences in the kinematical properties of the gas of the HH 80N core found to the east and to the west of IRS1: while the western gas has a constant velocity of $\sim12$~\kms, the eastern gas has a velocity gradient with the velocities getting blue-shifted to the east, and its line-width tends to broaden.

In order to study the kinematics of the N-bearing species, we obtained Position-Velocity plots (PV plots) of cuts along the major (P.A. = 122\arcdeg) and minor (P.A. = 32\arcdeg) axes of the HH 80N core, corresponding to the cuts A and B shown in Figure~\ref{ammonia_integrated}. The resulting PV plots for \ammonia~(1,1), \nhtd~(1$_{1,1}$--1$_{0,1}$) and \hntc~($1-0$) are shown in Figure~\ref{pvplots_n}. The PV plots along the major axis (see left panels in the figure)
display an inverse L-shaped morphology encompassing the SE and IRS1 dusty sources. \ammonia\ shows the most extended emission, while \nhtd\ and, specially, \hntc\ emission is clumpy being the last significantly less bright. The PV plots along the minor axis (see right panels in the figure) proves that \ammonia\ is the brightest N-bearing species at IRS1.

\section{Discussion}

\subsection{Discarding the molecular ring interpretation}

\citet{girart2001} and \citet{masque2009} interpreted the emission of several
molecular lines observed with BIMA in the HH 80N core as a ring-like structure with $6 \times 10^4$~AU of radius seen edge-on falling onto a protostar located at its center. This interpretation was based on the
morphology of the PV plots of the CS molecule, obtained along cuts similar  to the A and B cuts
presented in this paper. This morphology consists of an ellipse for the PV plot along the A direction and a double peak at same offset for the PV plot along the B direction. This is expected for the PV plots obtained with a flattened structure corresponding to a thin ring seen edge on, where the A direction represents the major axis and the B direction represents the minor axis of this structure. The N-bearing molecules presented in this paper are
expected to trace regions closer to the protostar than the species observed with
BIMA, for which models predict strong depletion at the low temperatures and high densities of these regions. 
Thus, the N-bearing molecules can help to determine the kinematical information of the inner parts of the core missed by CS, which would trace an outer shell.
To compare
the kinematics of CS with that of N-bearing species, we first convolved the \ammonia~(1,1), \nhtd~(1$_{1,1}$--1$_{0,1}$) and \hntc~(1--0) maps with a 2D Gaussian to obtain the same
synthesized beam as the BIMA CS (2--1) maps of \citet{masque2009}
($15.6\arcsec \times 7.1\arcsec$; P. A. = 3.0\arcdeg). We then obtained PV
plots of cuts along A and B directions (see Fig.~\ref{ammonia_integrated})
over the convolved maps. Note, however, that the cut along the major axis is
shifted 4\arcsec\ to the south with respect to the cut A presented in
\citet{masque2009}. The new cut traces better the emission of \ammonia, \nhtd\ and
\hntc\ along the major axis of the core. The resulting PV plots of these species superimposed over those of CS are shown in Figure~\ref{pvplots_cs}.

If the HH 80N core kinematics derived from CS \citep{girart2001,masque2009} corresponds to inside-out collapse with free fall motions, the gas closer to the protostar traced by the N-bearing molecules is expected to present larger velocities than the gas of the outer shells traced by CS. However, the velocity range of the gas traced by the \nhtd\ and
\hntc\ emission is similar or smaller than the velocity range of the gas traced by
CS. Thus, the gas motions detected with CS do not correspond to inside-out gravitational collapse. Furthermore, the appearance of the PV plots of \nhtd\ and \hntc\ along the major axis of the core is clearly asymmetric and has an L-shaped morphology, differing from the elliptical morphology expected for a contracting ring (or disk). 

The PV plots of the figure also show an extension and shape of \ammonia~(1,1) halfway between those of CS and \nhtd. Since \ammonia\ is expected to trace regions of the core with gas densities between those traced by \nhtd\ and those traced by CS, the kinematics of both gas regimes must contribute in the \ammonia\ behavior. Possibly, the diffuse gas in the core with moderate density mostly traced by CS ($n \sim10^4$~\cmt) includes larger velocity gradients than those seen with \nhtd\ and \hntc. This could produce a morphology reminiscent of a collapsing ring in the PV plots of CS, whose appearance is different from that of the PV plots of \nhtd\ and \hntc. 
Therefore, as a conclusion, the kinematic features of \nhtd\ and \hntc\ do not
match the features of a collapsing ring structure seen edge-on.

\subsection{Deuterium enrichment of the HH 80N core}

As seen before, the HH 80N core is composed of two condensations, SE and IRS1, having, at least the later, an embedded object. Both condensations are located where the \nhtd\ and \ammonia\ emission is brightest in the core. In this section we calculate $D_\mathrm{frac}$ (i.e. the ratio of \nhtd\ column density to \ammonia\ column density) over these two condensations as a measure of the degree of the deuterium enrichment. The degree of deuterium enrichment of chemical species in dense cores is found to increase with respect to the [D/H] elemental abundance ratio $\simeq10^{-5}$ \citep[estimated within 1~kpc of the Sun,][]{linsky2006} until the onset of star formation and to decrease afterward \citep{crapsi2005,emprechtinger2009}. The reason of the $D_\mathrm{frac}$ enhancement during the prestellar phase relies on the depletion of CO at the low temperatures ($< 20$~K) and high densities ($\simeq 10^6$~\cmt) found towards the central part of a prestellar core when it contracts. CO is the primary destroyer of H$_2$D$^+$, the main ion required to produce deuterated molecules via gas-phase reactions \citep[e.g.,][]{roberts2000, bacmann2003, pillai2007}, which is efficiently produced at low temperatures. Therefore, the freeze out of CO lets the [D/H] ratio propagate to other molecules via gas phase reactions with H$_2$D$^+$. Once a protostellar object is formed, $D_\mathrm{frac}$ decreases due to the internal heating of the core \citep{emprechtinger2009}. 
Deuterated ammonia, however, is also formed alternatively via grain-surface reactions with D atoms \citep{tielens1983}. This implies that, in addition to gas phase reactions, its gas abundance could be also affected by evaporation from dust grains caused by UV photons or outflow activity \citep[e.g.,][]{saito2000}.

In Table~\ref{classresults} we list the excitation temperature, main line optical depth and column density of \ammonia~(1,1) and \nhtd~(1$_{1,1}$--1$_{0,1}$), the rotational temperature of ammonia, and $D_\mathrm{frac}$ (calculation details are explained in the Table notes). As can be seen in the last column of Table~\ref{classresults}, the SE condensation has significantly higher $D_\mathrm{frac}$ than IRS1. Since the later condensation harbors an embedded protostar and shows signs of gas heating (\Trot\ $= 15.6$~K), the deuterium enrichment of the SE condensation could be a consequence of being 
in a prestellar evolutionary stage, earlier than that of IRS1. 
The only caveat to this trend is the presence of the UV radiation field coming from the Herbig Haro object HH 80N that impinges on the eastern part of the core \citep{masque2009}, as indicated by the moderate increase of \Trot, up to 13.5~K, observed toward this part. The UV photons could evaporate \nhtd\ from the grains, where it could be very abundant, and modify $D_{{\rm frac}}$. Indeed, the $D_\mathrm{frac}$ value around SE ($0.2-0.4$) is among the highest values derived in clustered star forming regions \citep[$0.1-0.8$,][]{busquet2010b,fontani2012}, where, in addition to the dense core evolution, $D_\mathrm{frac}$ can be locally affected by the interaction with UV radiation and/or molecular outflows of neighbouring protostars. However, the high visual extinction expected in the HH 80N core prevents the UV photons from penetrating inside \citep{masque2009}. Since \nhtd\ is found in the inner regions of the core likely far from the UV irradiation influence, the differences in $D_{{\rm frac}}$ probably represents different evolutionary stages between IRS1 and SE. 

If our derived $D_\mathrm{frac}$ values are not affected by the UV radiation field of HH 80N, then high gas densities must have been present in the HH 80N core for a long time to produce significant $D_\mathrm{frac}$ \citep[e.g. $\sim10^6$~yr,][]{howe1994}. Therefore, regardless the relative evolutionary state of the SE and IRS1 condensations, they were likely formed before the jet arrival.

\subsection{Dynamical interaction and triggered star formation?} 

In Sect. 3.1 we concluded that an infalling ring-like structure cannot explain the kinematics of the gas in the HH 80N core. Thus, other scenarios, such as that proposed by \citet{masque2011}, must be considered. These authors suggest that the HH 80N core constitutes a clump or a filament of gas, with several embedded condensations, possibly perturbed by the HH 80/81/80N outflow.

\begin{figure}[htbp]
\centering
\resizebox{0.65\textwidth}{!}{\includegraphics{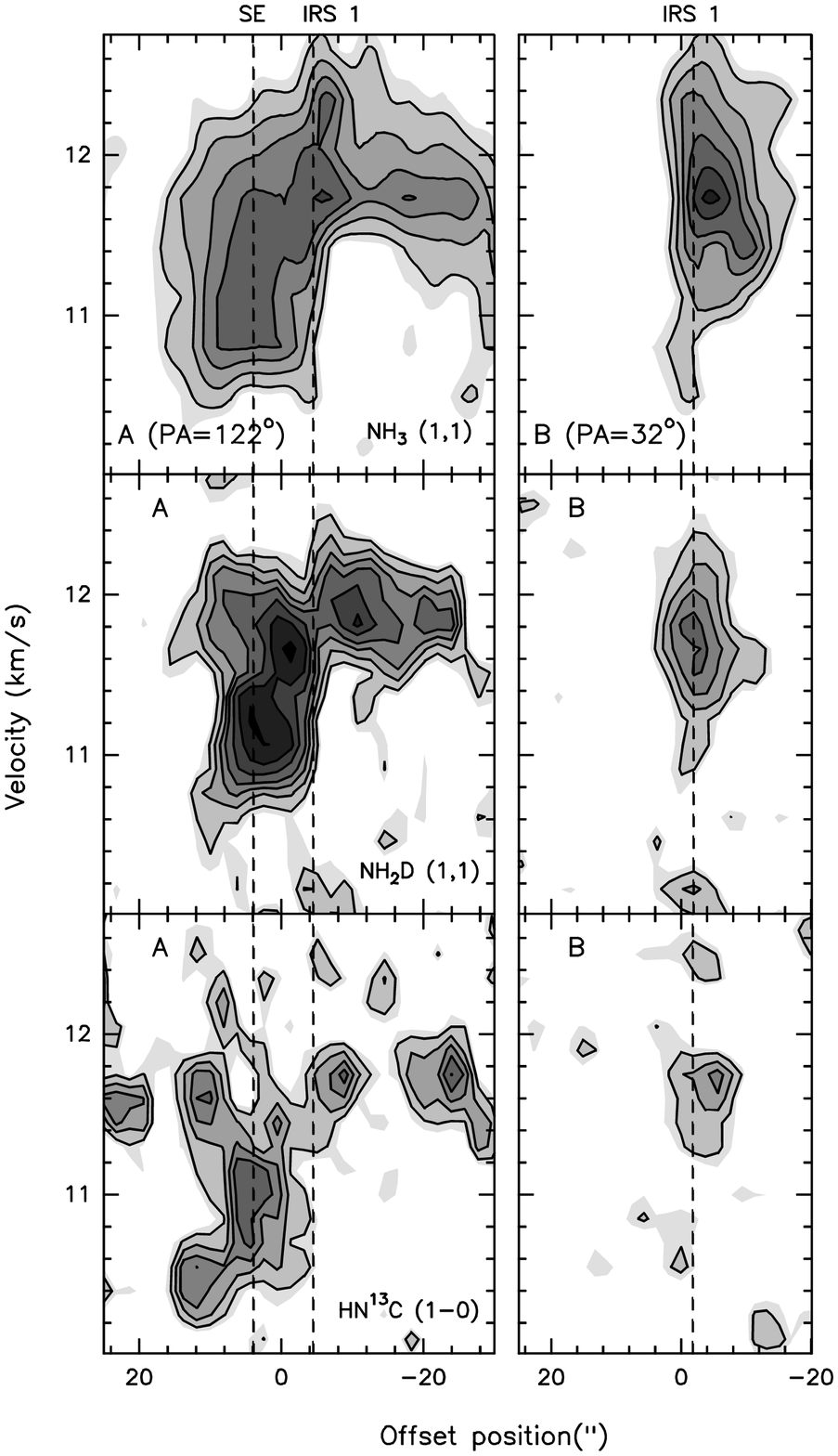}}
\caption{
PV plots of the \ammonia~(1,1) emission (top panels), \nhtd~(1,1) emission (middle panels) and \hntc~(1-0) emission (bottom panels) 
The panels on the left
represent the PV plots along the major axis of the core (PA = $122\arcdeg$, A direction in Fig.~\ref{ammonia_integrated}) and the panels on the right
represent the PV plots along the minor axis (PA = $32\arcdeg$, B direction in Fig.~\ref{ammonia_integrated}). 
The positive offsets of the PV plots
correspond to the southeast and northeast for the directions A and B, respectively. Contour levels are  3, 5, 7, 9, 12, 15 and 19 times  3.5~m\jpb\ for \ammonia, and 15~m\jpb\ for  \nhtd\ and  \hntc. The position offsets of IRS1 and SE are
marked with dashed lines. \label{pvplots_n}}
\end{figure}

\begin{figure}[htbp]
\centering
\resizebox{0.65\textwidth}{!}{\includegraphics{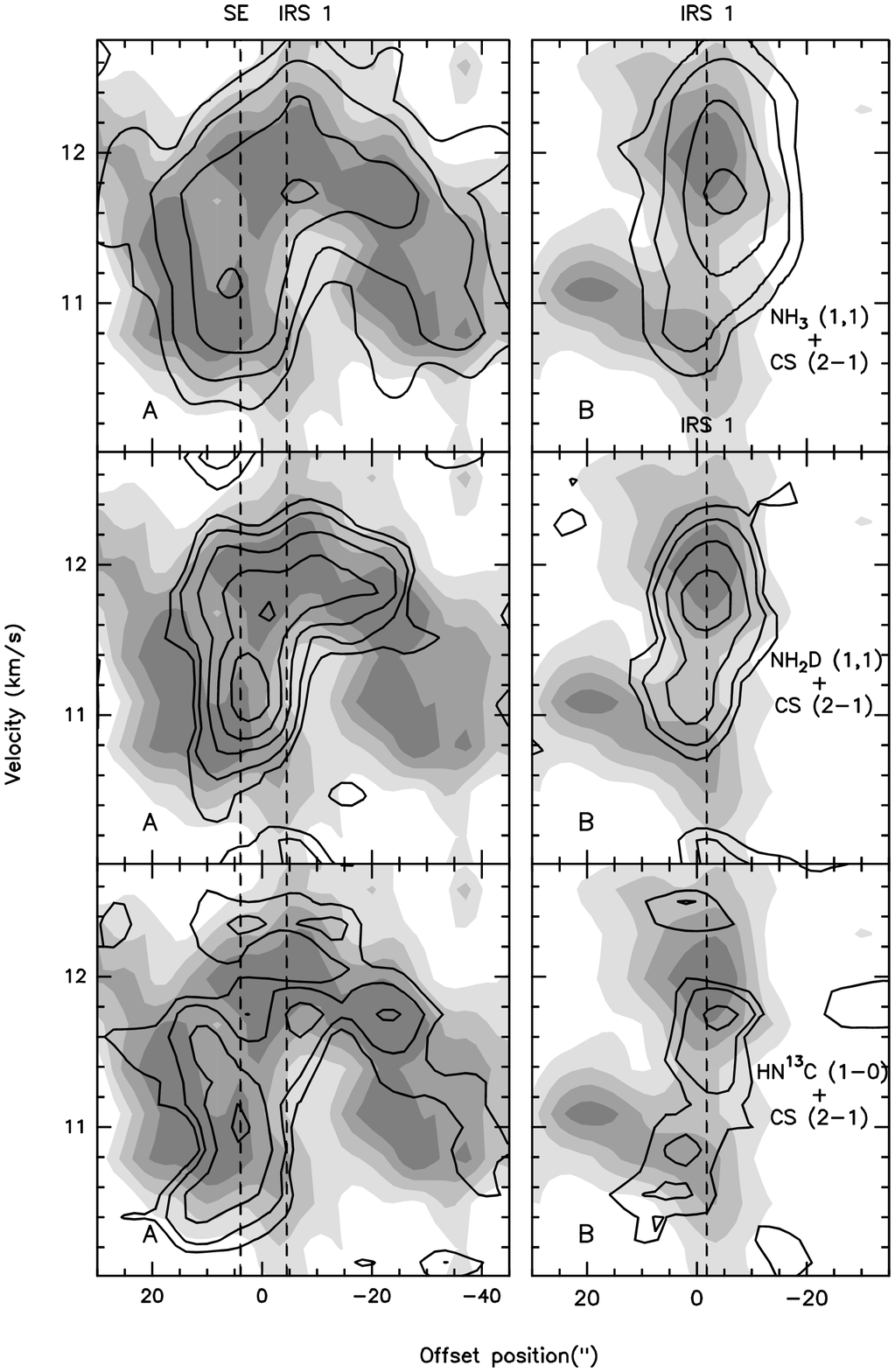}}
\caption{
PV plots of the \ammonia~(1,1) emission (top panels, contours), \nhtd~(1,1) (middle panels, contours) and \hntc~(1-0) emission (bottom panels, contours) superimposed over the PV plots of the CS (2-1) emission (gray scale) observed with BIMA \citep{masque2009}. The panels on the left
represent the PV plots along the major axis of the core (PA = $122\arcdeg$, A direction in Fig.~\ref{ammonia_integrated}) and the panels on the right
represent the PV plots along the minor axis (PA = $32\arcdeg$, B direction in Fig.~\ref{ammonia_integrated}). 
The positive offsets of the PV plots
correspond to the southeast and northeast for the directions A and B, respectively. The PV
plots of \ammonia~(1,1), \nhtd~(1,1) and \hntc\ (1-0) have been convolved with a Gaussian in order to obtain a resolution equivalent to the CS (2-1) BIMA observations (15.6\arcsec$ \times $7.1\arcsec, P.A. =
$3\arcdeg$). The contour levels are  3, 6, 12, 20 and 30 times 5~m\jpb\ for \ammonia, 20~m\jpb\ for \nhtd\ and 13~m\jpb\ for \hntc. The position offsets of IRS1 and SE are
marked with dashed lines. \label{pvplots_cs}}
\end{figure}

The bow-shock of the HH 80/81/80N outflow travels in the vicinity of the HH 80N core but does not impact against it as indicated by the lack of shock signatures. First, the linewidths observed in the core ($\sim1$~\kms) are significantly narrower than the linewidths derived in shocked regions \citep[6-7~\kms,][]{bachiller1995}. Second, the SiO emission, an excellent tracer of strong shocks in molecular gas, is absent in the HH 80N core \citep{masque2009}. However, the HH 80N core is located within the blue-shifted lobe of a large scale bipolar CO outflow found by \citet{benedettini2004}. These authors interpreted the outflow as a large portion of the molecular cloud set in motion by the HH 80/81/80N jet. Indeed, there is a velocity gradient in the HH 80N core, from IRS1 to SE, with the velocities in the eastern part of the core blue-shifted with respect to the central part (e.g. see Fig.~\ref{moments}). \citet{benedettini2004} found several CO knots along the jet axis. One of these knots (L), with a velocity of 10.1~\kms, is located $2'$ east of the HH 80N core. A seen in the PV plots along the major axis of the core (see Figs.~\ref{pvplots_n} and \ref{pvplots_cs}), the position and velocity of the L knot match the trend of the velocity gradient mentioned above, which suggest some dynamical connection of the eastern part of core with the outflow.

The velocity gradient inferred from the PV plots ($\sim5$~\kms~pc$^{-1}$) is of small magnitude, similar to the velocity gradients found in low-mass star-forming cores \citep[e.g.,][]{goodman1993, belloche2002, chen2007}. According to the authors, these gradients could be due to either rotation or slight outflow interaction with dense surrounding gas. Hence, we cannot discern between the outflow interaction with the HH 80N core or other gas motions as the origin of the velocity gradient found in the core. A similar scenario was found for the molecular cloud located ahead of HH 2: although it lacks of strong shock signatures, part of the cloud is possibly being driven out by the powerful winds from the VLA 1 protostar \citep{girart2005}. Thus, despite the lack of strong evidence for a dynamical perturbation of the HH 80N core by the HH 80/81/80N outflow, we still consider the possibility that part of the core is affected by the outflow.

In the case that outflow dynamical interaction is occurring, it could induce local instabilities in the HH 80N core that may trigger a
star--formation episode in the HH 80N region as presumably has happened in other regions (e.g. 1551: \citealt{yokogawa2003},
Serpens: \citealt{graves2010,duartecabral2010,duartecabral2011}). If a perturbed condensation becomes massive enough to induce gravity to overcome the internal pressure, then it could collapse. However, the PV plots along the major axis of Figs.~\ref{pvplots_n} and \ref{pvplots_cs} show that IRS1 is located at the limiting position between the region of the eastern velocity gradient and the rest of the core that moves at \vel\ of 12~\kms. In addition, because we deal with projected distances, IRS1 could be even further away from the outflow. Thus, the possible perturbation of the IRS1 condensation by the HH 80/81/80N outflow is unclear. On the other hand, the SE condensation falls in the region associated with the eastern velocity gradient and, hence, it could be located close to the interface of the presumable interaction. Moreover, \citet{masque2009} prove that the 'surface' of the eastern part of the core is affected by UV radiation from the HH 80N shock and, hence, there is a spatial association between this part of the core and the HH 80/81/80N outflow. However, in the previous section we found that the SE condensation seems to be in an earlier evolutionary stage than that of IRS1. This trend is opposite to the expected if the outflow has induced the star formation observed in the core, since its perturbation would affect first the SE condensation, the closest one to the outflow and, hence, SE should be the most evolved one. Therefore, the outflow triggered star formation in IRS1 is unlikely.

\section{Conclusions}

We present a kinematical study of the HH 80N core using nitrogen bearing species, aimed at determining the motions of the high density gas of the core ($n$~\gap~$10^5$~\cmt) missed by other molecular species previously observed in the region. In addition, we derived the
deuteration fraction in the core. Our main conclusions can be summarized as
follows: 

\begin{enumerate} 

\item The integrated emission of the N-bearing molecules (\ammonia, \nhtd\ and \hntc)  shows an elongated morphology similar (but more compact) to that of the species observed with BIMA (e.g. CS, SO, \hcom) or to the dark lane seen in the Spitzer 8~\mum\ image. Inspection of the zero, first and second-order moment maps of the N-bearing molecules reveals that most of the molecular emission arises from the eastern part of the HH 80N core, which is shifted $>0.5$~\kms\ in velocity and whose lines are broader than in the western part. Despite the lack of shock signatures in the core, these kinematical features suggest some perturbation in the eastern 
part, given the nearby location of the HH 80/81/80N outflow.

\item The morphology and kinematics of the high density regions of the core, revealed by the PV plots of the N-bearing species, rule out the previous interpretation as an infalling molecular ring-like structure with a radius of $6 \times 10^4$~AU inferred from BIMA observations of CS and other molecular species \citep{girart2001,masque2009}.

\item Assuming that the high extinction expected in the HH 80N core prevents the UV photons from the nearby HH 80N shock to influence the deuterium chemistry, the presence of \nhtd\ in the core implies that the SE and IRS1 condensations were likely present before the outflow reached the HH 80N core. Besides, the comparison of the degree of deuterium enrichment between the IRS1 and SE condensations suggests that the SE condensation is prestellar and less evolved than IRS1, consistent with the presence of an embedded protostar in the IRS1 condensation \citep{masque2011}. This trend is opposite to the expected if the outflow has induced the star formation observed in the core, since SE should be then the most evolved one. Therefore, the outflow triggered star formation in IRS1 is unlikely.

\end{enumerate} 

\acknowledgments

We thank Jos\'e F. G\'omez for his advice in some aspects
of the data reduction. We are grateful to Aina Palau for the fruitful discussion on the \nhtd\ chemistry.
J.M.M., J.M.G., G.A., M.O. and R.E. are supported by the Spanish MICINN AYA2008-06189-C03 and AYA2011-30228-C03 grants (co-funded with FEDER funds). J.M.G. is supported by the Catalan AGAUR 2009SGR1172 grant.

\bibliography{ammonia}


\begin{table}[ht]
\begin{center}
\caption{Interferometric observations. \label{liniesobservades}}
\vspace{0.5cm}
\begin{tabular}{lr@{.}lr@{~$\times$~}lr@{.}lcr@{.}lc}
\hline
\hline
Transition &  \multicolumn{2}{c}{$\nu$} & \multicolumn{2}{c}{Beam} & \multicolumn{2}{c}{PA} &$rms$ & \multicolumn{2}{c}{$\Delta v_{ch}$} & Telescope \\
          &    \multicolumn{2}{c}{(GHz)}  &\multicolumn{2}{c}{(\arcsec$\times$\arcsec)} & \multicolumn{2}{c}{($\arcdeg$)} & (mJy~beam$^{-1}$)&\multicolumn{2}{c}{(km~s$^{-1}$)}&   \\              
\hline
NH$_3$~(1,1)            & 23&6944955  &  6.4&  4.4  & 17&2          &4   &0&31 & VLA\\
NH$_3$~(2,2)            & 23&7226336  &  6.2& 4.3 & 18&3           &5   &0&31 & VLA \\
NH$_2$D~(1,1)     & 85&926263  &7.1&2.9 &10&0   & 20 &0&15 & PdBI \\
HN$^{13}$C~(1--0)  & 87&090859& 6.9&2.9 &10&9     & 23 &0&15 & PdBI \\
\hline
\end{tabular}
\end{center}
\end{table}

\begin{landscape}
\begin{table}[ht]
\footnotesize
\begin{center}
\vspace{-1.7cm}
\caption{Summary of the results of the \ammonia\ and \nhtd\ analysis. \label{classresults}}
\begin{tabular}{p{1cm}cccccccccc}
\hline
\hline
 &                 &  &  \multicolumn{3}{c}{\ammonia}  &    \multicolumn{3}{c}{\nhtd} &  &  \\ \cmidrule(r){4-6} \cmidrule(r){7-9}
    & \multicolumn{2}{c}{Coordinates$^\mathrm{a}$}    & \Tex~(1,1)$^\mathrm{b}$  &  $\tau_\mathrm{1,1,m}^\mathrm{c}$ & $N^\mathrm{d}$ & \Tex~(1,1)$^\mathrm{b}$  &$\tau_\mathrm{1,1,m}^\mathrm{c}$ & $N^\mathrm{e}$&\Trot~(NH$_3$)$^\mathrm{f}$&$D^\mathrm{g}_\mathrm{frac}$\\ \cline{2-3}
 Position        &    $\alpha$~(J2000) & $\delta$~(J2000) &(K) &  &(10$^{14}$~\cmd) & (K)& & (10$^{14}$~\cmd) &  (K) & \\  
\hline
SE$^\mathrm{h}$   & $18^\mathrm{h}19^\mathrm{m}18\fs52$      & $-20\arcdeg41'52\farcs9$& $6.8 \pm 0.8$  &  $1.8 \pm  0.4$    & $6.1-10.3$   & $5.6 \pm 0.6$  & $0.5 \pm 0.3$ & $2.1 \pm 0.3$ & $13.5 \pm 2.5$ & $0.2-0.4$ \\
Peak$^\mathrm{i}$   & $18^\mathrm{h}19^\mathrm{m}18\fs14$  &    $-20\arcdeg41'52\farcs0$  & $7.8 \pm 1.0$& $1.5 \pm 0.4$ & $15.5-5.8^\mathrm{j}$ & $5.3 \pm 0.5$ & $0.9 \pm 0.3$ &  $3.1 \pm 0.3$ & $10-15.8^\mathrm{j}$  & $0.2-0.6^\mathrm{j}$\\
IRS1   &  $18^\mathrm{h}19^\mathrm{m}17\fs82$   & $-20\arcdeg41'50\farcs2$ & $12.0 \pm 1.6$ & $0.6 \pm  0.4$  &  $3.8-4.9$  & $11.7 \pm 0.6$ &  $\leq0.2$   & $\leq0.5$ &$15.6 \pm 2.3$        & $\leq 0.1$ \\
\hline
\end{tabular}
\end{center}
\vspace{-0.3cm}
\textbf{Notes.} \\
$^\mathrm{a}$~Coordinates of the center of the box of about a beamsize used to average the \ammonia\ and \nhtd\ spectra.\\
$^\mathrm{b}$~Excitation temperature derived from the output parameters of a fit to the hyperfine structure of the (1,1) transition using CLASS. \\
$^\mathrm{c}$~Main line opacity derived from the fits to the hyperfine structure.\\
$^\mathrm{d}$~Beam-averaged column density of \ammonia\ obtained following the procedures given in \citet{sepulveda2011}. Upper limit is obtained from
$$
\left [ \frac{N(\mathrm{NH_3})}{\mathrm{cm^{-2}}} \right ] =1.58 \times 10^{13}\frac{e^{1.14/T_\mathrm{ex}}+1}{e^{1.14/T_\mathrm{ex}}-1}Q(T_\mathrm{rot})\tau_\mathrm{1,1,m} \left [ \frac{\Delta v}{\mathrm{km~s^{-1}}}  \right ] 
$$
where $T_\mathrm{ex}$ is obtained from the output CLASS parameter, $A \tau_\mathrm{1,1,m}$, where $A = f (J(T_\mathrm{ex})-J(T_\mathrm{bg})$, $J(T) = (e^{h\nu/kT}-1)^{-1}$ is the intensity in units of temperature,  $T_\mathrm{bg}$ is the background temperature, $f$ is the filling factor, which is assumed to be equal to 1, and $Q(T_\mathrm{rot})$ is the equipartition function at \Trot. 
We assumed that the rotational energy levels of the molecule are populated at $T_\mathrm{rot}$ under LTE conditions and, for $Q(T_\mathrm{rot})$, we only include the three lower rotational levels, which yields to  $Q(T_\mathrm{rot}) = (e^{23.4/T_\mathrm{rot}}+5e^{-41.5/T_\mathrm{rot}}+14e^{-101.2/T_\mathrm{rot}}+3)/3$.

The lower limit is obtained assuming $T_\mathrm{ex} \gg T_\mathrm{bg}$, which makes the explicit dependence on  $T_\mathrm{ex}$ disappear and the beam averaged column density is proportional to the output CLASS parameter, $A \tau_\mathrm{1,1,m}$, reducing to
$$
\left [ \frac{N(\mathrm{NH_3})}{\mathrm{cm^{-2}}} \right ] =2.78 \times 10^{13} \left  [ \frac{A \tau_\mathrm{1,1,m}}{\mathrm{K}}  \right ] Q(T_\mathrm{rot}) \left [ \frac{\Delta v}{\mathrm{km~s^{-1}}}  \right ] 
$$
 \\
$^\mathrm{e}$~Beam-averaged column density of \nhtd\ obtained with the equation 
$$
\left [ \frac{N(\mathrm{NH_2D})}{\mathrm{cm^{-2}}} \right ] =1.63 \times 10^{11} e^{20.68/T_\mathrm{ex}} Q(T_\mathrm{ex}) J(T_\mathrm{ex}) \tau_\mathrm{1,1,m} \left [ \frac{\Delta v}{\mathrm{km~s^{-1}}}  \right ]  
$$
similarly as for the \ammonia. In this case, we assumed LTE conditions with all the levels populated with the same $T_\mathrm{ex}$, and $Q(T_\mathrm{ex}) = 3.90 + 0.75 T_\mathrm{ex}^{3/2}$ \citep[Cologne Database for Molecular Spectroscopy,][]{muller2001}. As $N(\mathrm{NH_2D})$ uncertainties, we adopted the relative uncertainties of $A \tau_\mathrm{1,1,m}$ though the error in the determination of $N$ could be somewhat larger. \\
$^\mathrm{f}$~Rotational temperature derived using the \ammonia~(1,1) and \ammonia~(2,2) transitions following the expression (4) of \citet{ho1983} assuming the same $T_\mathrm{ex}$ for the \ammonia (1,1) and \ammonia~(2,2) lines. The \ammonia~(2,2) parameters have been obtained by fitting a single Gaussian to the spectra using CLASS.\\
$^\mathrm{g}$~$D_\mathrm{frac} = N$(\nhtd)/$N$(\ammonia).  \\
$^\mathrm{h}$~The box encloses the \ammonia~(2,2) eastern spot. \\
$^\mathrm{i}$~The box encloses the \ammonia~(1,1) peak.\\
$^\mathrm{j}$~The upper limit is obtained adopting the $3\sigma$ rms level of the maps for the \ammonia~(2,2) line intensity. The lower limit is obtained adopting a $T_\mathrm{rot}$ of 10~K, which is typically the lower temperature expected for dense cores.  
\end{table}
\end{landscape}

\end{document}